# Electro-optic dual-comb interferometry over 40-nm bandwidth

VICENTE DURÁN, PETER A. ANDREKSON, AND VÍCTOR TORRES-COMPANY*

Department of Microtechnology and Nanoscience (MC2), Chalmers University of Technology, SE 41296, Gothenburg, Sweden
*Corresponding author: torresv@chalmers.se

**Dual-comb interferometry is a measurement technique that uses two laser frequency combs to retrieve complex spectra in a line-by-line basis. This technique can be implemented with electro-optic frequency combs, offering intrinsic mutual coherence, high acquisition speed and flexible repetition-rate operation. A challenge with the operation of this kind of frequency comb in dual-comb interferometry is its limited optical bandwidth. Here, we use coherent spectral broadening and demonstrate electro-optic dual-comb interferometry over the entire telecommunications C band (200 lines covering ~ 40 nm, measured within 10 microseconds at 100 signal-to-noise ratio per spectral line). These results offer new prospects for electro-optic dual-comb interferometry as a suitable technology for high-speed broadband metrology, for example in optical coherence tomography or coherent Raman microscopy.**

Dual-comb spectroscopy (or, more generally, dual-comb interferometry) is a measurement technique capable to resolve the individual components of an optical frequency comb [1]. In a dual-comb spectrometer, the complex response of a sample is encoded on the spectral lines of a frequency comb acting as a probe. The basic idea for the implementation is using a second comb with slightly different line spacing as a local oscillator (LO). Compared to state-of-the-art Fourier-transform spectrometers, dual-comb systems provide superior frequency resolution without significant loss in sensitivity, as has been demonstrated in diverse spectroscopic applications in the last decade [2]. Additionally, the ability of dual-comb spectrometers to perform fast spectral complex measurements has been utilized in applications different from spectroscopy, such as optical arbitrary waveform characterization [3], distance measurement at long ranges with nanometer resolution [4] or hyperspectral imaging of vibrational transitions through coherent Raman spectroscopy [5].

Fiber or Ti:Sa modelocked lasers have been extensively employed for dual-comb spectroscopy. These laser sources provide millions of lines with a relative spacing of ~10-100 MHz, which is more than adequate for molecular spectroscopy. In these dual-comb systems, the offset in line spacing between combs fixes the minimum acquisition time to the millisecond scale. In practice, multiple spectra must be coherently averaged to increase the signal-to-noise-ratio, often limiting the measurement speed to ~1s [6]. Long acquisition times demand a high degree of mutual optical phase coherence between the combs, which must be carefully phase-locked through active feedback stabilization. If free-running femtosecond lasers are used instead, long-term operation requires adaptive dual-comb implementations [7] or real-time signal processing techniques [8].

Dual-comb spectroscopy can alternatively be implemented with electro-optic frequency combs. An inherent advantage of electro-optic generators for dual-comb interferometry is that a single continuous-wave (CW) laser can feed both combs simultaneously, thus achieving optical phase locking by default [2]. In addition, since electro-optic combs are generated without the need of an optical cavity, the line spacing can be easily tuned and reach values substantially larger than those provided by standard modelocked laser oscillators [9]. Several electro-optic dual-comb systems have been reported for fast, high-sensitive spectroscopy in the near infrared region [10-13], the measurement of optical arbitrary waveforms [3,14], or the characterization of optical communication components [3]. The number of lines in electro-optic dual-comb spectrometers rarely exceeds 100, although additional lines can be obtained by optical gating [13]. A common drawback in all previously reported electro-optic dual-comb spectrometers is that the operational bandwidth is only a few nm wide at best.

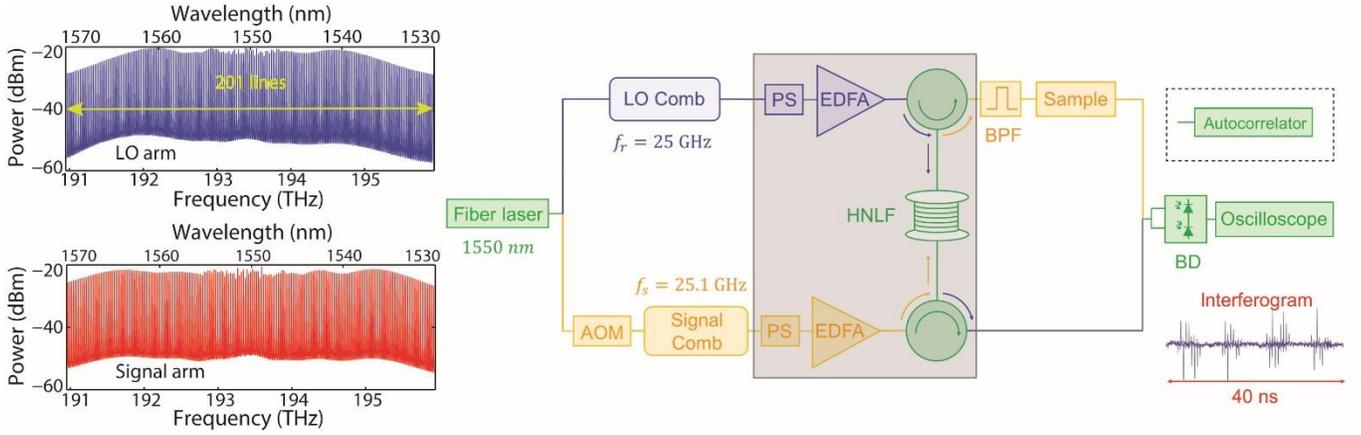

Fig. 1. Schematic of the experimental setup (see details in the text). The yellow line indicates the path follow by the signal, the blue line the path corresponding to the LO and the green line the common path for both arms. The dashed area comprises the part of the system used for achieving the coherent nonlinear broadening. On the left, flat-topped broadened spectra at the output of the HNLF for both interferometer arms.

In this paper, we report a dual-comb spectrometer implemented with electro-optic combs that operates over the entire telecommunications C-band. The central concept lies in realizing external broadening of the parent combs in a highly nonlinear fiber [13]. A key distinctive feature of our system is that it is implemented with high-performance 25 GHz parent combs whose transform-limited pulse durations lie in the sub-picosecond time scale [14], thus enabling efficient nonlinear broadening [15-16]. The offset in repetition rate frequencies and number of lines are carefully synthesized to fill the whole Nyquist measurement band (0-12.5 GHz) and still provide a single-shot acquisition speed in the sub-microsecond time scale. The bandwidth and number of lines is roughly four times larger than our previous results [14] and, to the best of our knowledge, the largest demonstrated for any electro-optic dual-comb system. The combination of bandwidth and measurement speed offers new possibilities of electro-optic combs for metrological applications that benefit from the large line spacing available in this platform such as Raman spectroscopy [5] or optical coherence tomography [17].

Our experimental setup is shown in Fig. 1. It includes two electro-optic combs (the signal and the LO) operating at 25 and 25.1 GHz repetition rate (thus the offset between combs is $\delta f = 100$ MHz). Each comb generator is composed of an intensity modulator followed by a pair of phase modulators. This arrangement provides a flat-topped spectrum composed of 55 lines at -10 dB [14]. Both combs are fed by a CW laser centered at 1550 nm, with a linewidth of 100 kHz. Each comb is spectrally broadened in a single HNLF as shall be described below. The interference between the signal and the LO is measured by a balanced detector. This acquired signal is digitized by a real-time oscilloscope with 33 GHz bandwidth. A temporal trace is recorded during 40 μs at a sampling rate of 50 GS/s. In the frequency domain, the interference between the sample and the LO leads to a multi-heterodyne detection process, since each line of the signal comb beats with all the lines from the LO. The resulting beat notes are distributed in sets (Nyquist zones) along the radio-frequency spectrum, leading to downconversion of the optical frequencies. To avoid aliasing in the RF spectrum (i.e., to ensure that the downconversion is unambiguous), an acousto-optic modulator (AOM) is included in the signal arm to shift the frequency of the CW laser [3]. Following [14], the frequency shift introduced by the AOM is chosen to be commensurate to $\delta f$ (concretely $f_{AOM} = \delta f/4 = 25$ MHz). Since $f_{AOM}$ is smaller than $\delta f$, its inverse fixes the minimum duration of a single interferogram ($T = 1/f_{AOM} = 40$ ns, see right lower inset in Fig. 1). The number of optical comb lines is filtered by a bandpass filter (BPF) so that after downconversion they fall within the first Nyquist zone (from dc to 12.5 GHz). For each interferogram, the spectral complex amplitude of the sample is recovered through an FFT routine.

The coherent spectral broadening of the combs is conducted in a counterpropagating configuration using a normal dispersion highly nonlinear fiber (HNLF) [13]. The HNLF is 100 m long, has a nonlinear coefficient of 11 (W km)$^{-1}$, a dispersion parameter $D = -1.03$ ps/nm/km at 1550 nm and a dispersion slope $S = 0.005$ ps/nm$^2$/km. The parent combs have a transform-limited duration of $\sim 660$ fs. In each interferometer arm, a programmable pulse shaper (PS) is used to reshape the pulses coming from the combs to a Gaussian profile of ~1.5 ps duration. The pulses are then amplified by an erbium-doped fiber amplifier (EDFA). The average powers at the input of the HNLF are 29.9 and 32.2 dBm for the reference and signal trains, respectively. The Gaussian profile of the input pulses, combined with the propagation along the HNLF, allows the pulses to enter into the optical wave-breaking regime. This process helps in reshaping the spectral broadening to a relatively flat optical spectrum [18]. At each fiber end, the comb is separated with the aid of a circulator. As can be observed in Fig. 1, the broadened spectra for the signal and the LO combs are very similar. They comprise 201 spectral lines that span over the C band. The power variation along this bandwidth is lower than 9 dB.

Our dual-comb spectrometer requires a calibration process to work [14]. This process implies the measurement of a trace when no spectroscopic sample is inserted in the signal arm. In this way, one retrieves the default relative complex amplitude between the two interferometer arms. We use this reference measurement to calculate the frequency-domain signal-to-noise ratio (SNR) for each comb line, $SNR_f(v)$. In spectroscopy, this magnitude is defined as $SNR_f(v) = |S(v)|/\sigma_v$, where $v$ is the optical frequency, $|S(v)|$ the amplitude of the power spectrum and $\sigma_v$ its standard deviation [19]. Figure 2(a) shows the measured $SNR_f(v)$ and the error in the retrieved spectral phase $\epsilon_\varphi(v)$ for a single-shot acquisition time (i.e., at a refresh rate of 25 MHz) and when 25 waveforms are averaged (effective refresh rate of 1 MHz). Assuming that $S(v)$ has the same uncertainty in each quadrature, the inverse of $SNR_f(v)$ should give $\epsilon_\varphi(v)$ [19]. This relationship is roughly fulfilled by our results. As can be observed in the plot, the outer comb lines (those with the lowest power) are clearly affected by higher phase noise. However, this effect can be compensated for by coherent averaging. On the other hand, and irrespective of the predominant noise source in the detection process, the spectral SNR scales as a $\sqrt{N}/M$, being $M$ the number of comb lines and $N$ the number of averaged interferograms [2]. A manner of quantifying the performance of a dual-comb interferometer is by calculating $SNR_f \times M$, where $SNR_f$ is the result of averaging $SNR_f(\nu)$ over all lines. The value of this product for an acquisition time of 1 s constitutes a figure of merit usually employed in dual-comb spectroscopy [2,19]. Figure 2(b) shows the values of $SNR_f \times M$ versus $N$ (i.e. as a function of the effective refresh rate) for $M = 201$ lines (red dots). In the same plot (blue dots), we also include the values of the above product excluding the nonlinear broadening stage (dashed box in Fig. 1). In this case, $M = 55$, and the resultant dual-comb system is very similar to the one used in [14], except for the fact that we use here a CW laser with slightly broader linewidth. This difference explains that both curves do not scale as $\sqrt{N}$ (see the discontinuous line calculated for $M = 55$). The performance of our electro-optic dual-comb system is better for $M = 55$, especially at high effective refresh rates, by less than a factor of 3 dB. The decrease in $SNR_f$ for $M = 201$ is likely due to spontaneous emission noise in the amplifiers and a decrease of power per line. This degradation calls for realizing coherent averaging to maintain the same SNR, thus decreasing the effective refresh rate. At 100 kHz (i.e., averaging $N$=250 waveforms), $SNR_f \times M \sim 2\times10^4$ for $M = 201$. This performance is comparable to that achieved for similar values of $N$ for previously reported electro-optic dual-comb interferometers (see, for instance, [12]).

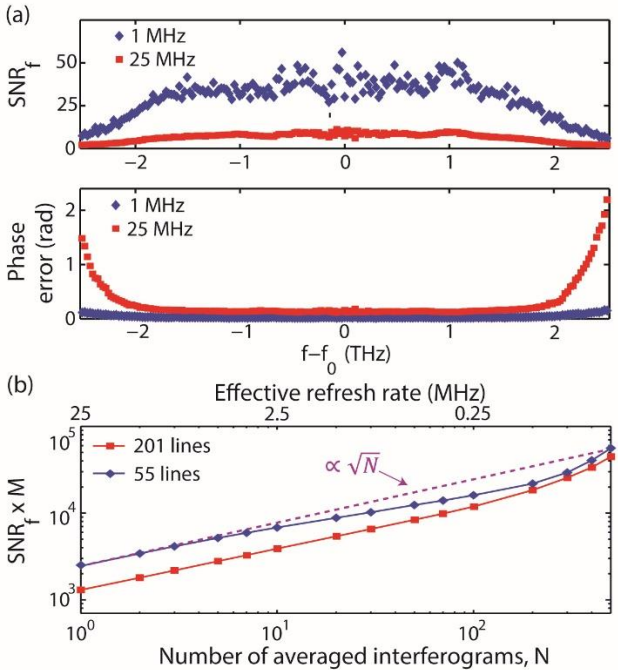

Fig. 2. (a) Spectral SNR and the corresponding spectral phase error $\epsilon_\varphi$ for 25 MHz (single-shot acquisition) and 1 MHz. In the latter case, the maximum value of $\epsilon_\varphi$ is 150 mrad. (b) Plot of $SNR_f \times M$ when coherent averaging is performed for two configurations of our system.

We evaluate the performance of our system with two relevant measurements: a complex arbitrary waveform programmed with the aid of a line-by-line pulse shaper [20] and the measurement of the dispersion introduced by low dispersive samples. For the first example, we impart onto the sample comb spectrum a phase profile corresponding to a sinusoidal phase function with two abrupt phase changes of $\pi/2$, see the blue curve in Fig. 3(a). The recorded trace has a duration of 40 $\mu s$. By averaging 250 waveforms (i.e., at an effective refresh rate of 100 kHz), $SNR_f$ was increased to 70. Thus, we reach a value for the spectral phase error, averaged over the whole C band, $\epsilon_\varphi = 14$ mrad (equivalent to an optical path difference of $\sim\lambda/440$). This value is comparable to that achieved by a previous configuration of our system with $M = 55$ at $\sim$1 MHz [14]. Figure 3(b) shows the intensity pulse profile calculated from the retrieved spectral complex amplitude (amplitude and phase). As can be observed, the pulses emerging from the HNLF are reshaped into a temporal pulse formed by an irregular sequence of peaks, which span over the entire comb period. The measurement of this kind of temporal signals constitutes a challenge due to the possible overlapping between consecutives pulses [20]. However, it can be easily retrieved with high fidelity in the sub-millisecond scale with the dual-comb approach.

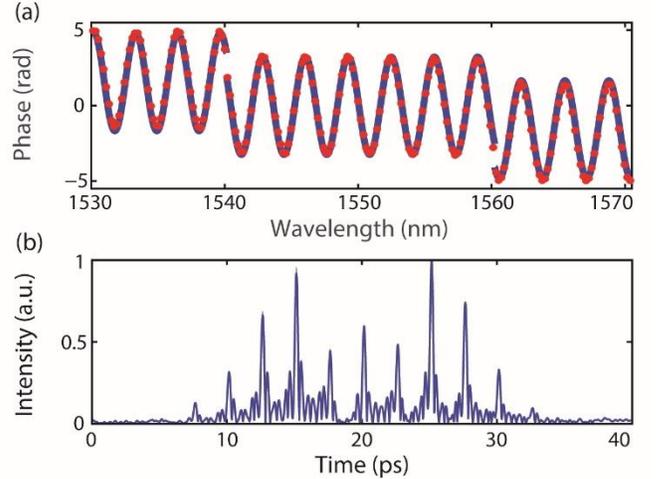

Fig. 3. (a) Phase profile imparted onto the signal spectrum by the pulse shaper (blue curve) and spectral phases obtained from a single interferogram at 100 kHz (red points). (b) Reconstructed intensity profile built from the electro-optic dual-comb measurement.

The broad bandwidth achieved by our system can be alternatively used to measure the dispersion introduced by a variety of spectroscopic samples. As a preliminary experiment for our interferometer, a spool of 19.92 m of single-mode fiber (SMF-28) is inserted in the signal arm. We measure 16 sequences of interferograms with the oscilloscope, each one of 10 $\mu$s of duration. From each sequence, we retrieve the quadratic spectral phase introduced by the fiber along the C band (5 THz). Figure 4(a) shows the mean value of $\varphi$ for each comb line. In the lower inset, the standard deviation for the central lines is represented by the area delimited by the two broken lines. The mean phase deviation of $\varphi$ along 5 THz is just $\sim 0.3\%$ of the measured phase range. A fit of $\varphi$ (blue curve) allows us to calculate the quadratic dispersion $\beta_2$ of the fiber, which results to $\beta_2 = -21.93 \pm 0.02$ ps$^2$/km. This corresponds to a dispersion parameter $D$ of 17.21 ps /(km nm), in good agreement with the value provided by the manufacturer.

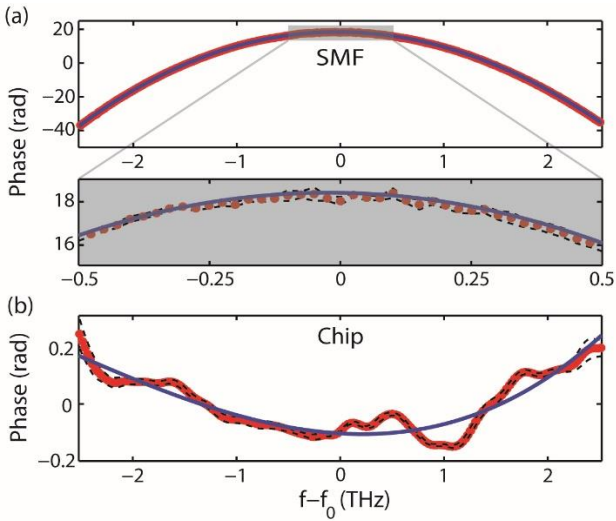

Fig4. (a) Recovered spectral phase corresponding to a spool of 20m of single-mode fiber over 5 THz of bandwidth. (b) Recovered spectral phase for a photonic chip waveguide of 8 mm of length with normal dispersion. In both cases, the phase standard deviation is represented by broken lines and a polynomial fit is included in blue.

The accuracy of the above result points out that broadband complex spectra can be precisely measured with our dual-comb spectrometer. This ability is especially suitable for measuring the low-dispersion introduced by a photonic waveguide. As an example, we consider a dual-core silicon nitride waveguide (manufactured using commercially available TriPleX™ technology [21]) with a total length $L$=8 mm. From the cross-section geometry the waveguide is expected to introduce a small amount of normal dispersion in the 1.5 $\mu$m band. Prior to the dispersion measurement with the dual-comb system, we measure the dispersion of the waveguide using a pulse shaper and an intensity autocorrelator as follows. We only use the signal arm of the setup shown in Fig. 1. A broader comb (comprising 300 lines within 60 nm of bandwidth) is achieved by increasing the power sent into the HNLF. The balanced detector is replaced by a commercial autocorrelator. The previous pulse shaper is now used as a programmable filter. The power going into the waveguide is -4dBm and we do not observe nonlinear broadening into the chip. The method for measuring the dispersion is conducted in two steps. In the first one, no sample is present. A correcting spectral phase $\Psi(\nu)$ is programmed onto the pulse shaper to generate transform-limited pulses at the input of the autocorrelator. When the chip is placed in the signal arm, the dispersion broadens the pulses, and their signature becomes evident in the autocorrelation trace. With the aid of the pulse shaper, we add an additional quadratic phase term to compensate for the distortion introduced by the dispersion of the waveguide. The programmed dispersion onto the pulse shaper reversed in sign provides an estimation of the dispersion introduced by the silicon nitride waveguide. With this method, we find a normal dispersion $\beta_2 = 338 \pm 40 \ ps^2/km$. Figure 4(b) shows the spectral phase $\varphi(\nu)$ measured for the waveguide using our dual-comb spectrometer using the same procedure and combs as in Fig. 4(a). Here, 6 independent traces of 40 μs are recorded and each one processed to retrieve the spectral phase of the sample. Note that since the measurement is performed in a line-by-line manner, higher-order dispersion terms can be fitted. The main values of $\varphi$ are shown as red dots, and their standard deviation are represented by broken lines. The mean phase deviation of $\varphi$ is around 3% of the measured phase range. By means of a polynomial fit, the quadratic dispersion parameter can be found. The outer comb lines are removed from the fit, giving $\beta_2 = 364 \pm 9 \ ps^2/km$. This value is fully compatible with the measurement provided by the pulse shaping method.

In conclusion, we have presented an electro-optic dual-comb spectrometer that overcomes the modest bandwidth in previously reported electro-optic dual-comb systems. Our spectrometer operates over the entire C band at refresh rates of ~100 KHz. We have provided examples regarding the measurement of complex waveforms approaching a 100 % duty cycle and of low-dispersive spectroscopy samples. The combination of high-measurement speed, large comb line spacing and bandwidth here reported will be of interest for applications of electro-optic frequency combs beyond molecular absorption spectroscopy, e.g. in Raman microscopy, or optical coherence tomography.

We acknowledge funding from the Swedish Research Council and the Marie Curie actions Intra European Fellowship (PIEF-GA-2013-625121), Career Integration grant (PCIG13-GA-2013-618285) and grant agreement ERC-2011-AdG - 291618 PSOPA.